\documentclass[final,5p,times,twocolumn]{elsarticle}

\usepackage{amssymb}
\usepackage{amsmath}

\journal{Optics and Lasers in Engineering}

\begin{document}

\begin{frontmatter}



\title{All-Optical Brillouin Random Number Generator}


\author[a,b]{Artem Mukhamedyanov}
\author[a,b,c]{Evgeny Andrianov}
\author[a,b,c]{Alexander Zyablovsky}

\affiliation[a]{organization={Moscow Center for Advanced Studies},
            addressline={20 Kulakova Str},
            postcode={141700}, 
            state={Moscow},
            country={Russia}}
\affiliation[b]{organization={Dukhov Research Institute of Automatics (VNIIA)},
            addressline={22 Sushchevskaya},
            postcode={127055}, 
            state={Moscow},
            country={Russia}}
\affiliation[c]{organization={Institute for Theoretical and Applied Electromagnetics},
            addressline={13 Izhorskaya},
            postcode={125412}, 
            state={Moscow},
            country={Russia}}

\begin{abstract}
We propose a model of binary random number generator (RNG) based on a Brillouin optomechanical system. The device uses a hard excitation mode in a Brillouin optomechanical system, where thermal noise induces spontaneous transitions between two stable states in the hard excitation mode. We demonstrate the existence of an amplitude criterion for observing these transitions and show that the probability distribution of their occurrence in the non-generating and generating states can be precisely controlled by the amplitude of an external pump wave. At the same time, the use of a low-intensity seed wave allows for the control of the transition times between states. We demonstrate that the proposed random number generator successfully passes the standard tests NIST SP 800-22. The obtained result opens a way for development of an all-optical integrated True RNG, generating a sequence of random bits with equal probability.
\end{abstract}


\end{frontmatter}



\section{Introduction}
\label{sec1}
Random number generators (RNGs) have significant applications in stochastic simulation \cite{3, 0} and Bayesian neural networks \cite{2, 2.1}. True RNGs can be used for the simulation of various stochastic physical \cite{breuer} and biological systems \cite{4, 5}. Modern stochastic methods realized on classical computers are often computationally slow and frequently rely on Monte Carlo method \cite{2.1, 3, 5.5}. This method involve calculating statistical properties, such as mean and dispersion, by calculating a numerous number of random samples.

Recent advances have led to the development of integrated photonic computing systems. These systems can perform operations, such as matrix multiplication or computing trained neural networks, to accelerate classical computations. Photonic computing is thus a promising technology for future hardware. Various photonic elements can serve as sources of stochastic signals \cite{6}, including vertical-cavity surface-emitting lasers (VCSELs) \cite{7}, semiconductor ring lasers \cite{8} and etc.

Several variants of optical random number generators, which are integrated into optical circuits, have been proposed \cite{opt_rng1, opt_rng2}. These systems use laser radiation noise or nonlinear effects as the basis for random number generation. However, such systems require digital signal processing after noise measuring to extract the bit stream. Currently, chaos \cite{opt_mech_chaos} and stochastic resonance \cite{opt_mech_stoch}, as well as noise-induced random transitions \cite{pt_symm}, are actively studied in optomechanical systems. Such induced chaos and spontaneous transitions in the optomechanical system might the basis for generating True RNG.

In this paper, we demonstrate a novel RNG concept based on a Brillouin optomechanical system. Brillouin lasers are promising candidates for nonlinear elements within photonic integrated circuits. We consider the Brillouin laser operating in the hard excitation mode \cite{hard_exc,fotiadi2020}. In this regime, thermal noise can lead to spontaneous transitions between the non-generating and generating states \cite{pt_symm}, which cause jump-like changes in the laser intensity. Such jumps in the signal can be considered as a sequence of binary numbers (zero and ones). We use analytical and numerical models to derive the criterion for observing such stochastic transitions. Based on this criterion, we define the laser parameters suitable for generating random signals. We demonstrate that using a scheme with two waves: a pump wave and a seed wave make it possible to control the generation rate and distribution of random numbers. We demonstrate that the proposed RNG based on the Brillouin laser successfully passes the standard tests NIST SP 800-22 \cite{NIST}. The obtained results open the way to creation of a binary RNG on-chip devices.

\section{Model}
\begin{figure}[t]
\centering\includegraphics[width=\linewidth]{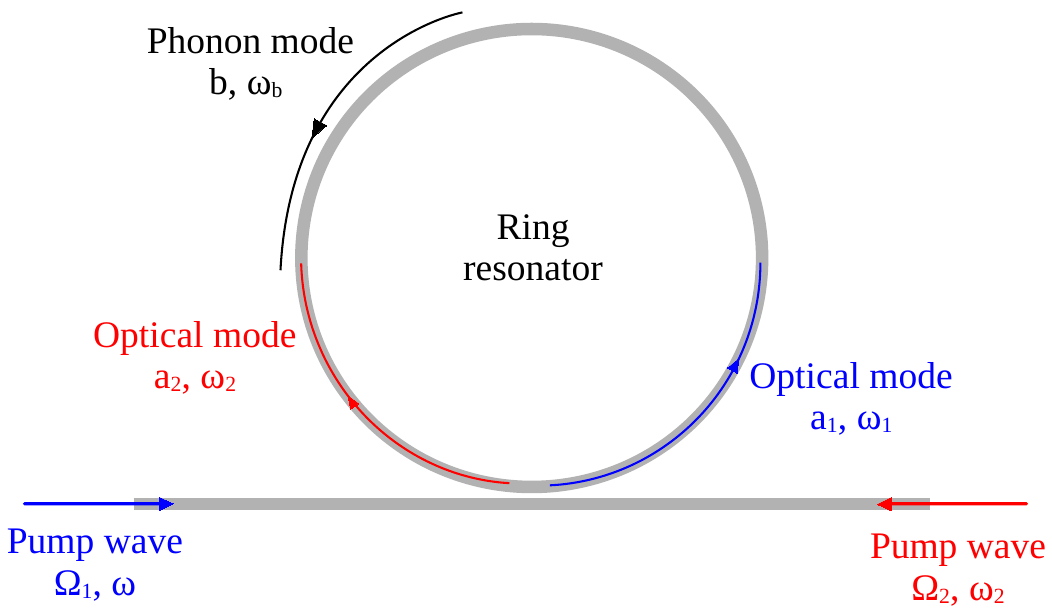}
\caption{Scheme of the system under consideration.}
\label{fig:1}
\end{figure}

We consider a Brillouin optomechanical system consisting of a ring resonator that is connected to an optical waveguide Fig.~\ref{fig:1}. We study the behavior of two optical modes in the ring resonator, which interact with each other via a phonon mode. The first optical mode with frequency $\omega_1$ is excited by the external pump wave with amplitude $\Omega_1$ and frequency $\omega$. The amplitude of the external pump wave is given as $\Omega_{1} = \sqrt{\frac{\kappa_{ex} P_{1}}{\hbar \omega}}$, where $\kappa_{ex}$ is the decay rate of input-cavity coupling, $P_{1}$ power of the corresponding optical pumps. The frequencies of the optical modes, $\omega_{1,2}$ ($\omega_1 > \omega_2$), are determined by the length of the ring resonator and the order of the optical mode in it. We consider the system where the frequency of the phonon mode coincides with the difference of the frequencies of the optical modes ($\omega_b = \omega_1 - \omega_2$). To describe the system, we use the optomechanical Hamiltonian \cite{1}:

\begin{equation}
\begin{gathered}
  \hat H = \hbar {\omega _1}\hat a_1^\dag {{\hat a}_1} + \hbar {\omega _2}\hat a_2^\dag {{\hat a}_2} + \hbar {\omega _b}{{\hat b}^\dag }\hat b +  \hfill \\
  \hbar \,g\left( {\hat a_1^\dag {{\hat a}_2}\hat b + {{\hat a}_1}\hat a_2^\dag {{\hat b}^\dag }} \right) + \hbar \Omega_{1} \left( {\hat a_1^\dag {e^{ - i\omega t}} + {{\hat a}_1}{e^{i\omega t}}} \right) + \hfill \\
  \hbar \Omega_{2} \left( {\hat a_2^\dag {e^{ - i\omega_2 t}} + {{\hat a}_2}{e^{i\omega_2 t}}} \right)
\end{gathered}
\label{eq:1}
\end{equation}
Here ${\hat a_{1,2}}$ and $\hat a_{1,2}^\dag $ are the bosonic annihilation and creation operators for the first and second optical modes, respectively. $\hat b$ and ${\hat b^\dag}$ are the bosonic annihilation and creation operators of the phonon mode. $g$ is the coupling strength between the optical modes and the phonon mode (Frohlich constant). The fifth term describes the pumping of the first optical mode by an external pump wave. $\Omega_1$ is proportional to the amplitude of the pump wave. The last term describes the excitation of the second optical mode by a low-intensity seed wave. This seed wave is used to control the states in the optomechanical system. $\Omega_2$ is proportional to the amplitude of the seed wave.

The system under consideration is an open quantum system. To describe relaxation processes and thermal noise in the optomechanical system, we use the Heisenberg-Langevin approach \cite{carmichael,gardiner,scully}. Within this framework, we derive the equations for the c-number amplitudes with noise of the optical and phonon modes that have the form \cite{subthreshold,hard_exc,pt_symm,thresh_control}:

\begin{equation}
\frac{{d{a_1}}}{{dt}} = \left( { - {\gamma _1} - i{\delta\omega _1}} \right){a_1} - ig{a_2}b - i\Omega_{1}
\label{eq:2}
\end{equation}

\begin{equation}
\frac{{d{a_2}}}{{dt}} = \left( { - {\gamma _2} - i{\Delta _2}} \right){a_2} - i{g}{a_1}{b^*} - i\Omega_{2} e^{-i(\delta\omega_{2} + \delta\omega)t}
\label{eq:3}
\end{equation}

\begin{equation}
\frac{{db}}{{dt}} = \left( { - {\gamma _b} - i{\Delta _b}} \right)b - i{g}{a_1}a_2^* + \xi(t)
\label{eq:4}
\end{equation}
Here ${a_{1,2}}$ and $b$ are the c-number amplitudes of the optical modes and the phonon mode, respectively. ${\gamma _{1,2}}$, ${\gamma _b}$ are the relaxation rates of the respective quantities. $\delta\omega_1 = \omega - \omega_1$ is the frequency detuning between the first optical mode and the pump wave. $\delta\omega_2 = \omega - \omega_2$ is the frequency detuning between the pump wave and the second optical mode. We consider that the frequency of the seed wave coincides with the one of the second optical mode. $\Omega_{1,2}$ are the amplitudes of the pump and seed waves, respectively. $\delta\omega = \dfrac{\omega_{b} \gamma_2 - \delta\omega_2 \gamma_b}{\gamma_2 + \gamma_b}$ is the frequency of generated phonons \cite{hard_exc}. $\Delta_2 = \delta\omega_2 + \delta\omega = \dfrac{\delta \omega_1}{\gamma_2 + \gamma_b} \gamma_{2}$
and $\Delta_{b} = \omega_b - \delta\omega = \dfrac{\delta \omega_1}{\gamma_2 + \gamma_b} \gamma_{b}$. $\xi \left( t \right)$ is the thermal noise that acts on the phonon mode, the correlation function of which is proportional to ${\gamma _b}\bar n$, where $\bar n$ is the average number of thermal phonons in the system \cite{hard_exc} (for details, see Appendix A).

\begin{figure*}[t]
\centering\includegraphics[width=1\linewidth]{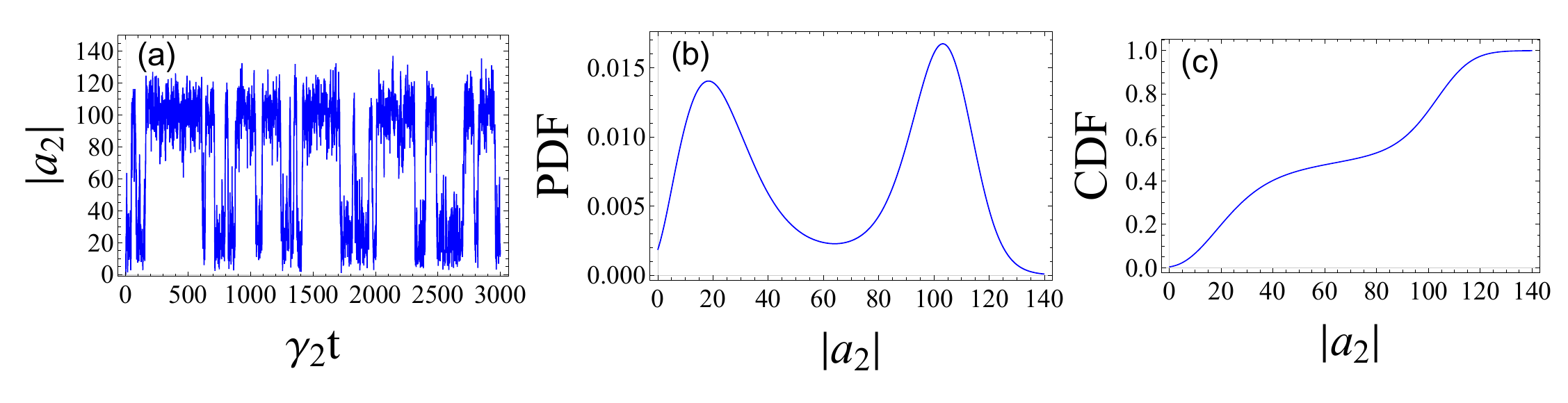}
\caption{Time dependence of the amplitude of the first optical mode (a), the probability density function (PDF) (b) and the cumulative distribution function (c). Here ${\gamma _1} = {\gamma _2} = 2\pi \cdot 191$ MHz, ${\gamma _b} = 2\pi \cdot 1.2$ GHz, $\delta {\omega _{\,1}} = 2\pi \cdot 1.58$ GHz , $\delta {\omega _2} = -2\pi \cdot 10.59$ GHz , ${\omega _b} = 2 \pi \cdot 12.17$ GHz, $g = 2\pi \cdot 15.9$ MHz , $\Omega_1 = 7.83 \cdot 10^{-1} \Omega_{th}$ and $\bar n = 513$. These parameters are close to the experimental ones \cite{linbo, sin, waveguide, strongcouple}.}
\label{fig:2}
\end{figure*}

\section{Criterion for Spontaneous Transitions}
In the system under consideration, there is a parameters region in which the hard excitation mode takes place \cite{hard_exc}, in which a jump-like increase in the laser intensity is observed with an increase in the pump wave intensity. The hard excitation mode is realized when ${\delta\omega_1}^2 > \gamma_1 (\gamma_2 + \gamma_b)$. In the case of the hard excitation mode, there exists a region of bistability with two stable states: generating and non-generating ones. The region of bistability is defined by the inequality $\Omega_{\mathrm{ex}} < |\Omega_1| < \Omega_{\mathrm{th}}$, where $\Omega_{\mathrm{ex}} = \dfrac{\sqrt{\gamma_b \gamma_2}}{|g|} \left| \delta\omega_1 \left( 1 +  \dfrac{\gamma_1}{\gamma_2 + \gamma_b} \right) \right|$ is the excitation amplitude and $\Omega_{\mathrm{th}} = \dfrac{\sqrt{\gamma_b \gamma_2}}{|g|}\sqrt{ \left( \gamma_1^2 + \delta\omega_1^2 \right) \left[ 1 + \left( \dfrac{\delta\omega_1}{\gamma_2 + \gamma_b} \right)^{\!2} \, \right] }$ is the threshold amplitude \cite{hard_exc}. It has been demonstrated that thermal noise can induce spontaneous transitions between these two states~\cite{pt_symm}.

From the analysis of the stationary generating solution of Eqs.~(\ref{eq:2})--(\ref{eq:4}), we can obtain the criterion for spontaneous transitions. To do this, we will consider the case where the seed wave amplitude is zero ($\Omega_2 = 0$). In this case, the generating solutions for the optical mode $a_2$ and the phonon mode $b$ are \cite{hard_exc}:

\begin{equation}
\left| a_2 \right|^2 = \frac{1}{|g|} \sqrt{\frac{\gamma_b}{\gamma_2}} \sqrt{ |\Omega_1|^2 - \Omega_{\mathrm{ex}}^2 } + \frac{\gamma_b}{\gamma_2} \frac{ \delta\omega_1 \Delta_2 - \gamma_1 \gamma_2 }{ |g|^2 }
\label{eq:5}
\end{equation}
\begin{equation}
\left| b \right|^2  = \frac{1}{|g|} \sqrt{\frac{\gamma_2}{\gamma_b}} \sqrt{ |\Omega_1|^2 - \Omega_{\mathrm{ex}}^2 } + \frac{ \delta\omega_1 \Delta_2 - \gamma_1 \gamma_2 }{ |g|^2 }
\label{eq:6}
\end{equation}
Eqns.~(\ref{eq:5})-(\ref{eq:6}) describe the Brillouin laser generation. When $\delta\omega_1 \Delta_2 > \gamma_1 \gamma_2$, we obtain the hard excitation mode where the intensities of the optical mode $a_2$ and the phonon mode $b$ change abruptly \cite{hard_exc}. Using the expression for $\Delta_2$, this condition can be rewritten as

\begin{equation}
|\delta\omega_1| > \sqrt{\gamma_1 (\gamma_2 + \gamma_b)}
\label{eq:7.0}
\end{equation}

When the external pump amplitude exceeds the excitation threshold, $|\Omega_1| > \Omega_{\mathrm{ex}}$, the intensity of the phonon mode jumps to the value \cite{hard_exc}:

\begin{equation}
J_b = \frac{\gamma_2}{\left|g\right|^2} \left( \frac{ \delta\omega_1^2 }{ \gamma_2 + \gamma_b } - \gamma_1 \right)
\label{eq:7}
\end{equation}

To observe spontaneous transitions between two states, the average thermal noise amplitude must be greater than or comparable to the magnitude of the phonon mode jump ($\bar{n} \ge J_b$). However, when the noise magnitude is greater than the jump magnitude, it becomes difficult to distinguish the non-generating state from the generating one.
Therefore, for observing jumps, it is optimal when $J_b \sim \bar{n}$. Using this estimation and the Eq.~(\ref{eq:7}), we derive the condition for the frequency detuning necessary for observing intensity jumps:

\begin{equation}
|\delta\omega_1| < \sqrt{ (\gamma_2 + \gamma_b) \left( \frac{g^2}{\gamma_2} \bar{n} + \gamma_1 \right) }
\label{eq:8}
\end{equation}
Thus, the hard excitation mode takes place when $|\delta\omega_1| > \sqrt{\gamma_1 (\gamma_2 + \gamma_b)}$. In this case, there is a region of parameters where spontaneous transitions are possible due to thermal noise. Combining the Eqns.~(\ref{eq:7.0}) and (\ref{eq:8}), we obtain that these transitions are observed when the following condition 

\begin{equation}
|g|^2 \bar{n} \sim \gamma_1 \gamma_2
\label{eq:8.2}
\end{equation}
is satisfied.

\section{Spontaneous transitions between the stable solutions}
\begin{figure}[t]
\centering\includegraphics[width=\linewidth]{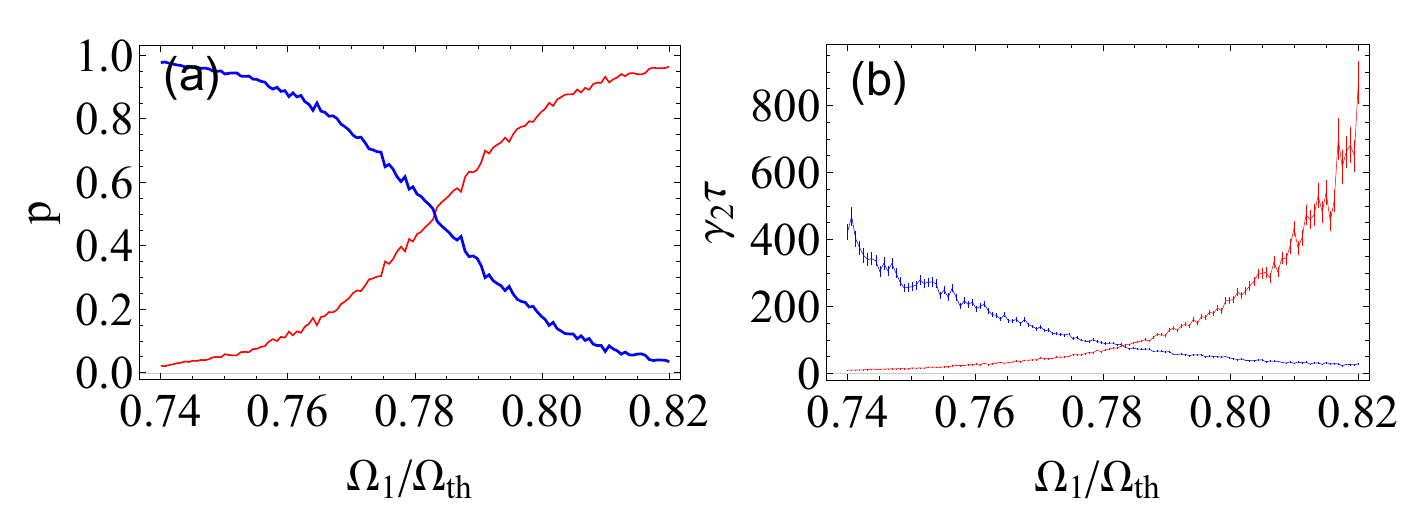}
\caption{(a) The dependence of the probabilities of being in the non-generating state, $p_{ng}$, (the blue line) and the generating state, $p_{g}$ (the red line). (b) The dependence of the average lifetimes of the non-generating (the blue line) and the generating (the red line) states on the pump amplitude $\Omega_1$ (a). The parameters are the same as in Fig.~\ref{fig:2}.} 
\label{fig:3}
\end{figure}

To quantify the transition probability between the two states induced by thermal noise, the system of Eqns.~(\ref{eq:2})-(\ref{eq:5}) was numerically simulated using Kloeden-Platen-Schurz algorithm \cite{breuer}. The simulation gives the time dependencies of each mode (Fig.~\ref{fig:2}(a)), from which the probability density function (PDF) can be derived. When there are transitions between two states, the probability density function has two maxima, one of which corresponds to the system being in a non-generating state, and the other to the system being in a generating state Fig.~\ref{fig:2}(b).
Using numerical integration, we can find a cumulative distribution function (CDF). The probability of the generating state is calculating from CDF. Using the cumulative distribution function, we calculate the probabilities of being in the generating, $p_g$, and non-generating, $p_{ng}$, states. By calculating the cumulative distribution function for different values of the pump amplitude, one can find the value of $\Omega_1$ at which the probabilities of being in non-generating and generating states are equal to each other ($p_{ng} = p_{g}$) Fig.~\ref{fig:3}. With our parameters, the corresponding value is $\Omega_1 = 7.83 \cdot 10^{-1}\Omega_{th}$ Fig.~\ref{fig:3}. From the temporal dynamics of the system Fig.~\ref{fig:2}(a), we can determine the average lifetimes of the system in the non-generating and generating states for different pump wave amplitudes Fig.~\ref{fig:3}(b). Hereinafter, $\tau_{g}$ is the average lifetime of the generating state, $\tau_{ng}$ is the average lifetime of the non-generating state. As expected, the lifetimes in the states are equal to each other when $p_{ng} = p_{g}$.

\begin{figure}[t]
\centering\includegraphics[width=\linewidth]{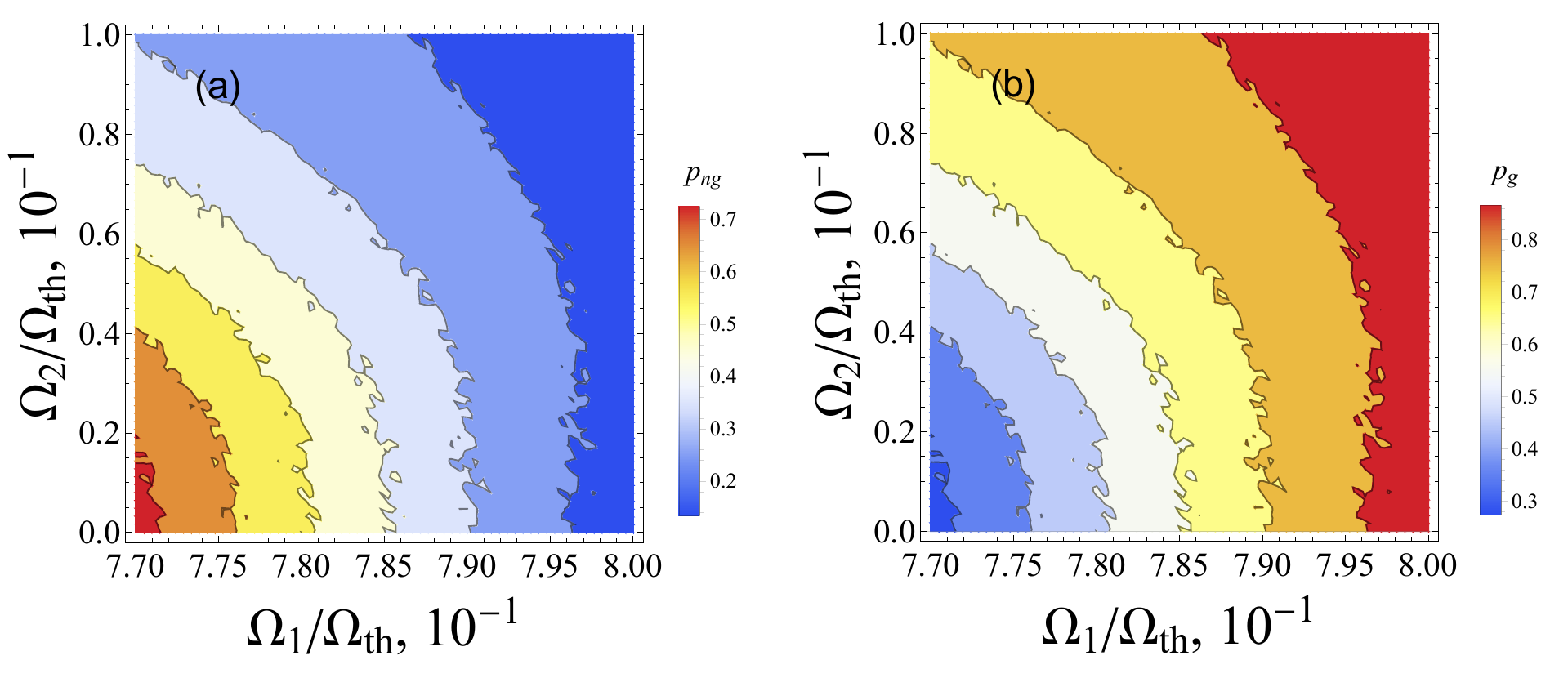}
\caption{The dependence of the probabilities of being in the non-generating state, $p_{ng}$, (a) and the generating state, $p_{g}$, (b) on the pump amplitude $\Omega_{1}$ and $\Omega_2$. The parameters are the same as in Fig.~\ref{fig:2}.} 
\label{fig:4}
\end{figure}

\begin{figure}[t]
\centering\includegraphics[width=\linewidth]{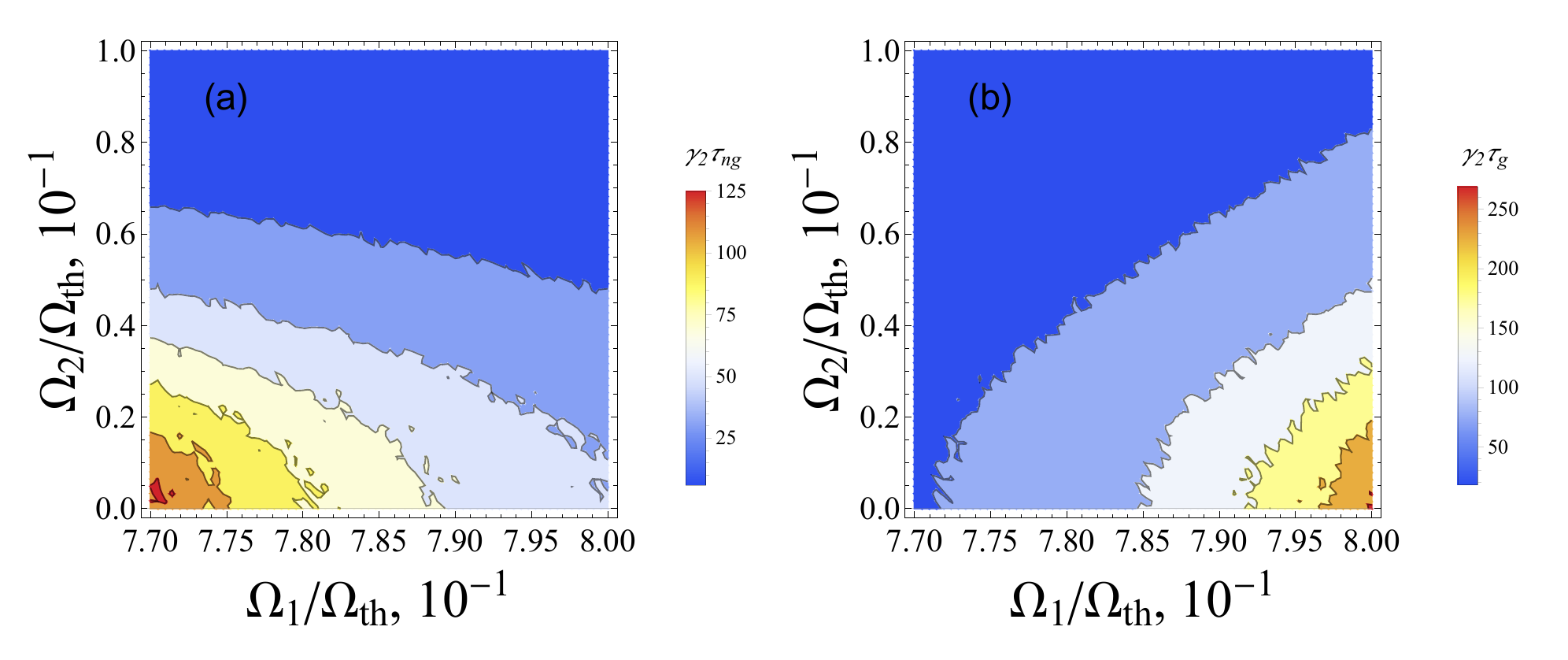}
\caption{The dependence of the average lifetimes of the non-generating state, $\tau_{ng}$, (a) and the generating, $\tau_{g}$, (b) on the pump amplitude $\Omega_{1}$ and $\Omega_2$. The parameters are the same as in Fig.~\ref{fig:2}. Calculation of the standard error of the obtained results is in \ref{app2}.} 
\label{fig:5}
\end{figure}

In the work \cite{thresh_control}, it has been demonstrated that the low-intensity seed wave can be used to control the stability and generation threshold of an optomechanical system operating in the hard excitation mode. Our numerical calculations show that using the seed wave allows for precise tuning of the probabilities $p_{ng}$ and $p_g$ Fig.~\ref{fig:4} and control the average lifetimes of both states ($\tau_{ng}$, $\tau_{g}$) Fig.~\ref{fig:5}.

\begin{figure}[t]
\centering\includegraphics[width=0.7\linewidth]{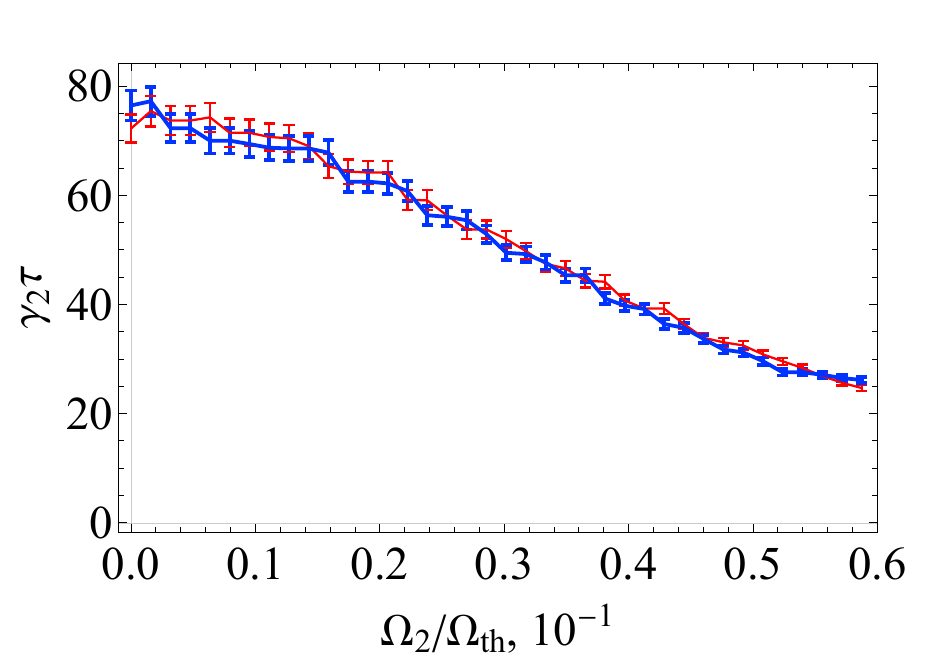}
\caption{The dependence of the average lifetimes of the generating state, $\tau_{g}$ (red solid line), and the non-generating, $\tau_{ng}$ (blue solid line), on the seed wave amplitude $\Omega_{2}$ when the probabilities are equal $p_{ng} = p_{g}$. The amplitude of the pump wave $\Omega_{1}$ is set so that the average lifetimes of the two states are equal to each other. The parameters are the same as in Fig.~\ref{fig:2}. The standard error is in range $0.49/\gamma_2 < \Delta\tau < 2.8/\gamma_2$.} 
\label{fig:51}
\end{figure}

Control of the lifetimes can be achieved with variation of the amplitude of the seed wave, $\Omega_2$, which changes the relative stability of the two states: generating and non-generating. As the amplitude of $\Omega_2$ increases, the stability of the non-generating state decreases, thereby increasing the probability of being in the generating state. From the Fig.~\ref{fig:4} and Fig.~\ref{fig:5} we can obtain a set of amplitudes of the pump $\Omega_1$ and the seed $\Omega_2$ waves where probabilities and average lifetimes are equal $p_{ng} = p_g$. The numerical simulation shows that the seed wave decreases the average lifetimes Fig.~\ref{fig:51} when probabilities are equal $p_{ng} = p_{g}$. By using the seed wave with an intensity much lower than the threshold value for the pump wave, it is possible to reduce lifetimes several times Fig.~\ref{fig:51}.

\section{Optomechanical Random Number Generator}

\begin{figure}[t]
\centering\includegraphics[width=1\linewidth]{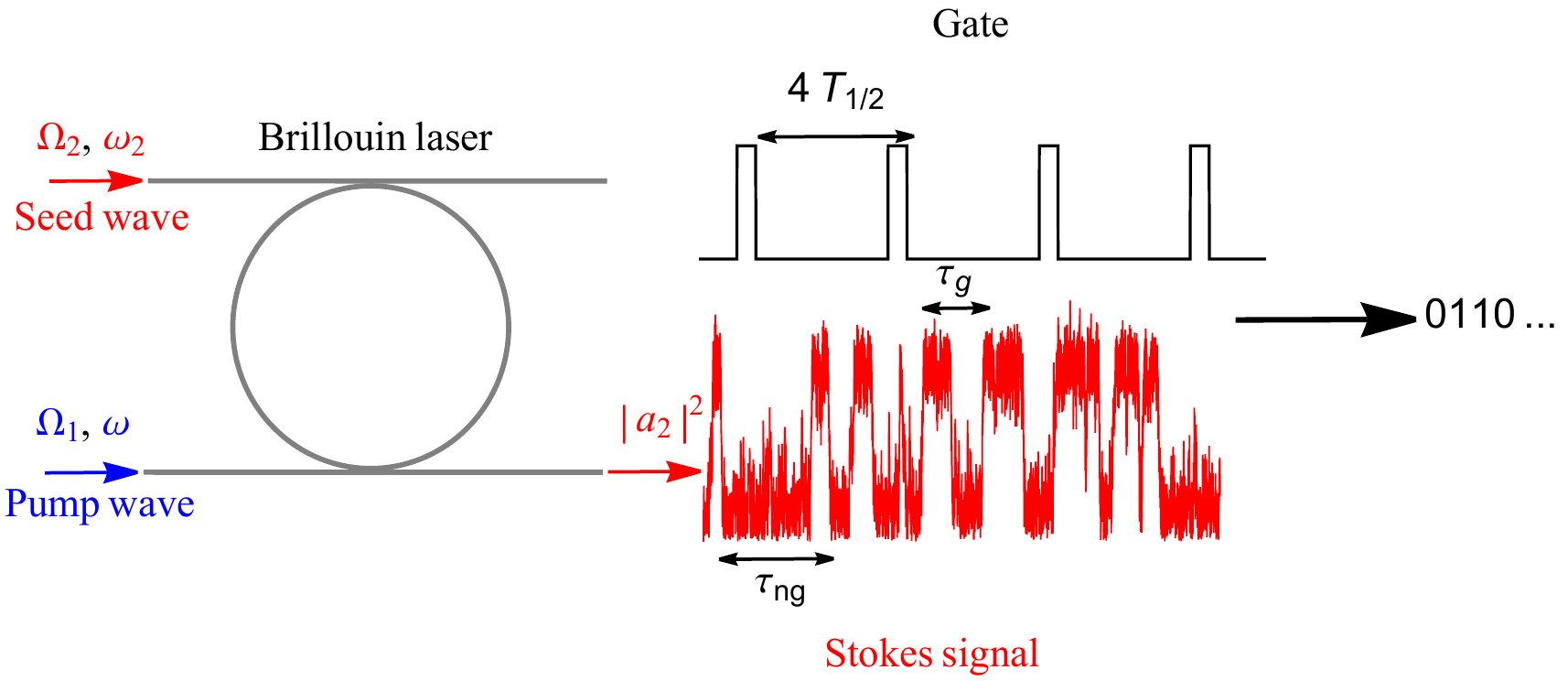}
\caption{The scheme of the optomechanical Random Number Generator.}
\label{fig:6}
\end{figure}

Based on the results for the behavior of the optomechanical system in the hard excitation mode, we propose an all-optical device that can be formally described as a coin-flip generator Fig.~\ref{fig:6}. This device can generate randomly a stream of bits ($0$, $1$). To verify that the process of the generating of random bits is true and there are no any correlations in data, we use standard tests NIST SP 800-22 \cite{NIST}.

To set up the device, firstly we estimate the pump amplitude of $\Omega_1$ where average lifetimes of the generating and the non-generating states ($\tau_{g}$ and $\tau_{ng}$) are equal to each other. Secondly, we estimate a boundary amplitude value of the second mode, $|a_{2bound}|$ \cite{9, 10}, which separated the generating and the non-generating states. 
When the amplitude of the second optical mode is greater than the boundary value ($|a_2| > |a_{2bound}|$), it means that the current state is the generating one and formally can be considered as $1$. In opposite case when $|a_2| < |a_{2bound}|$, the current state is the non-generating one which can be considered as $0$. Finally, we estimate the sampling frequency. In the case when the average lifetime of the two states are equal ($\tau_{g} = \tau_{ng} = \tau$) we can introduce a half-life time $T_{1/2} = \tau \log{2}$. Here the half-life time $T_{1/2}$ is the lifetime of the both states: the non-generating and the generating ones. NIST tests required a balanced bit stream with equal amount of $0$ and $1$. The balance of bit stream can be setup by the amount of $|a_{2bound}|$. Furthermore, some NIST tests to prove randomly generated data required frequent transitions between states. We can control the transition frequency by the amplitude of the external seed wave, $\Omega_2$, or by varying the sampling frequency. In our model, to observe more frequent transitions between $0$, $1$ and pass all NIST test, we use the next sampling frequency $f_s = \dfrac{1}{4T_{1/2}}$. At this sampling frequency, the generator passes all NIST tests (see Appendix \label{app3}). Using parameters from the Fig.~\ref{fig:2}, we obtain that the speed of bit generating is about $5$ Mb/s. 

\section{Conclusion}
In summary, we have investigated the Brillouin optomechanical system operating in the hard excitation mode as an all-optical device for true random number generation. Using thermal noise and spontaneous transitions between two stable states, we have shown that in such system the probability of being in the states and the average lifetimes of the states can be controlled by the external pump and seed waves. The explanation of such phenomena because of the relative stability of the generating and non-generating states depend on the external waves. This allows to work in the regime where the probabilities of being in the states are equal to each other ($p_{ng} = p_g = 0.5$). This device can be used as random number generator without any digital data processing and have possible realization on chip. It provides a random bit generation which passes all standard NIST tests with the rate of generation of about $5$ Mb/s. 

The obtained value of the generation rate is less than those that can be achieved in other methods \cite{trng1, trng2}. The generation rate of the proposed RNG can be increased by the increase of the relaxation rates of the modes. In particular, the relaxation rate of the phonon mode, $\gamma_b$, determines the noise amplitude and directly affects the frequency of transitions between states. With any combination of parameters, the transition time between the generating state and the non-generating one cannot be less than the lifetimes of photonic and phonon modes that are inversely proportional to $\gamma_{1,2}$ and $\gamma_b$. That is, there is a fundamental limit which is determined by the photon and phonon lifetimes. With the used system parameters, the generation rate is limited by the rate of photon relaxation and is approximately 100 Mbit/sec. An increase in the relaxation rates leads to an increase in the generation threshold, $\Omega_{th}$, which requires the use of large pumping power for the operation of the proposed RNG. Therefore, depending on the tasks of the optical computing circuit, optimization is required between the bit stream and the energy efficiency of the RNG.

The advantage of the generator we offer is that it can be implemented in photonic integrated circuits. Our system requires gate which divide the signal on to the bits. The gate is periodic function which pass or block optical signal and the relaxation rate of such gate must be in order with generation rate. It can be realized by using only optical elements such as saturable absorbers where relaxation rates are near the sampling frequency rate. It means that the system can be realized as all-optical random bit generator by using optical resonator and saturable absorber. The further signal can be used in photonic integrated circuit which is consist of random bits and it is fully optical. The creation of an all-optical random number generator could be important for the development of optical computing devices. For example, recently, the use of Brillouin lasers as nonlinear activation function in all-optical neural networks on a chip has been discussed \cite{slinkov2025all}. The random number generator we propose can be implemented on similar Brillouin laser structures, which will make it possible to combine several different elements of computing devices on a single chip.

\section*{Acknowledgments}
A.R.M., E.S.A. and A.A.Z. thank the foundation for the advancement of theoretical physics and mathematics “Basis”.

\appendix
\section{Stochastic equation simulation method}
\label{app1}
In our numerical simulation we model the following system of stochastic equation 
\begin{equation}
d{a_1}= \left(\left( { - {\gamma _1} - i{\delta\omega _1}} \right){a_1} - ig{a_2}b - i\Omega_{1}\right)dt
\label{eq:A1}
\end{equation}

\begin{equation}
{d{a_2}} = \left(-\gamma _2{a_2} - i{g}{a_1}{b^*} - i\Omega_{2} \right)dt
\label{eq:A2}
\end{equation}

\begin{equation}
\begin{gathered}
{db}= \left(\left( { - {\gamma _b} - i(\omega_b + \delta\omega_2)} \right)b - i{g}{a_1}a_2^* \right)dt \\
+ \sqrt{\gamma_b \bar n}\left(dW_1 + idW_2 \right)
\label{eq:A3}
\end{gathered}
\end{equation}
The thermal noise $\xi(t)$ is implemented as the formal derivative of a Wiener process $\xi(t) = \xi_{1}(t) + i\xi_{2}(t) = (dW_1+i dW_2)/dt$ with correlation \[\left \langle dW_{1,2}(t) dW_{1,2}(t') \right \rangle = \begin{cases}
\sigma^2 dt, & t = t' \\
0, & t \neq t'
\end{cases}\] and $\left \langle dW_{1}(t) dW_{2}(t') \right \rangle = 0$. For numerical scheme we define the Wiener increment as $\Delta W_{1,2} \sim \sigma \sqrt{\Delta t} N(0,1)$. Here $N(0,1)$ is the normal distribution,$\Delta t$ is the time step of numerical scheme and $\sigma = \sqrt{\gamma_b \bar n}$ is the square root of variance (amplitude of the thermal noise), which determined from noise correlation function $\left\langle {{{\xi }^{*} }\left( {t'} \right) \xi \left( t \right)} \right\rangle  = 2{\gamma _b}{\bar n}\,\delta \left( {t - t'} \right)$ and the mean of the noise $\left\langle {\xi \left( t \right)} \right\rangle  = 0$. Here $\gamma_b$ is relaxation rate of the phonon mode, $\bar n$ is the average number of thermal phonons in the system. The average number of thermal phonons are defined from the Bose-Einstein distribution. At the room temperature this is simplified to the classical approximation $\bar n \approx \hbar \omega_b / k_B T$ . Here $\omega_b$ is the frequency of the phonon mode, $T = 300$ K room temperature. Next we use Kloeden-Platen-Schurz scheme \cite{breuer} and discretitize the Eqns.~(\ref{eq:A1}) - (\ref{eq:A3}).

\section{States lifetime distribution and error}
\label{app2}

\begin{figure}[t]
\centering\includegraphics[width=0.7\linewidth]{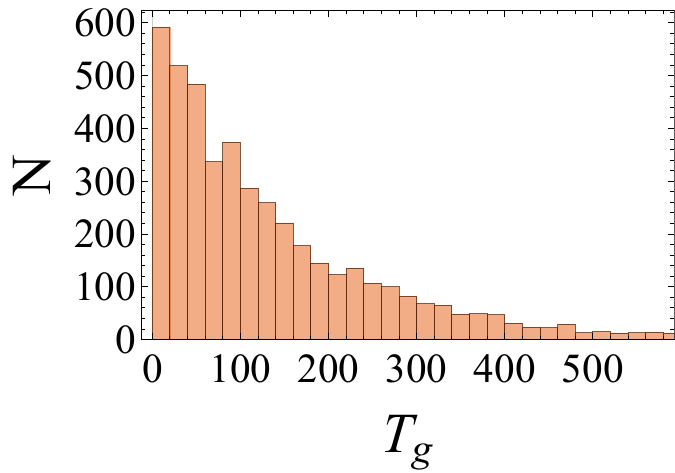}
\caption{The distribution of lifetimes for generating state. The pump amplitude is $\Omega_1 = 7.9 \cdot 10^{-1} \Omega_{th}$.}
\label{fig:B1}
\end{figure}

In our simulations, the state lifetimes $\tau_{g/ng}$ is defined as the average value over a set of a generated sample \[ \tau_{g, ng} = \frac{1}{N} \sum_{n = 1}^{N} T^{n}_{g, ng}\] Here $T^{n}_{g, ng}$ is the $n$-th value of the lifetime of the generating and the non-generating states. We note that the lifetimes follow an exponential distribution (Fig. \ref{fig:B1}), with the mean of the lifetime distribution equal to its square root of variance. The standard error for such distribution can be estimated as $\Delta \tau_{g/ng} = \tau_{g, ng}/\sqrt{N}$.

\begin{figure}[t]
\centering\includegraphics[width=\linewidth]{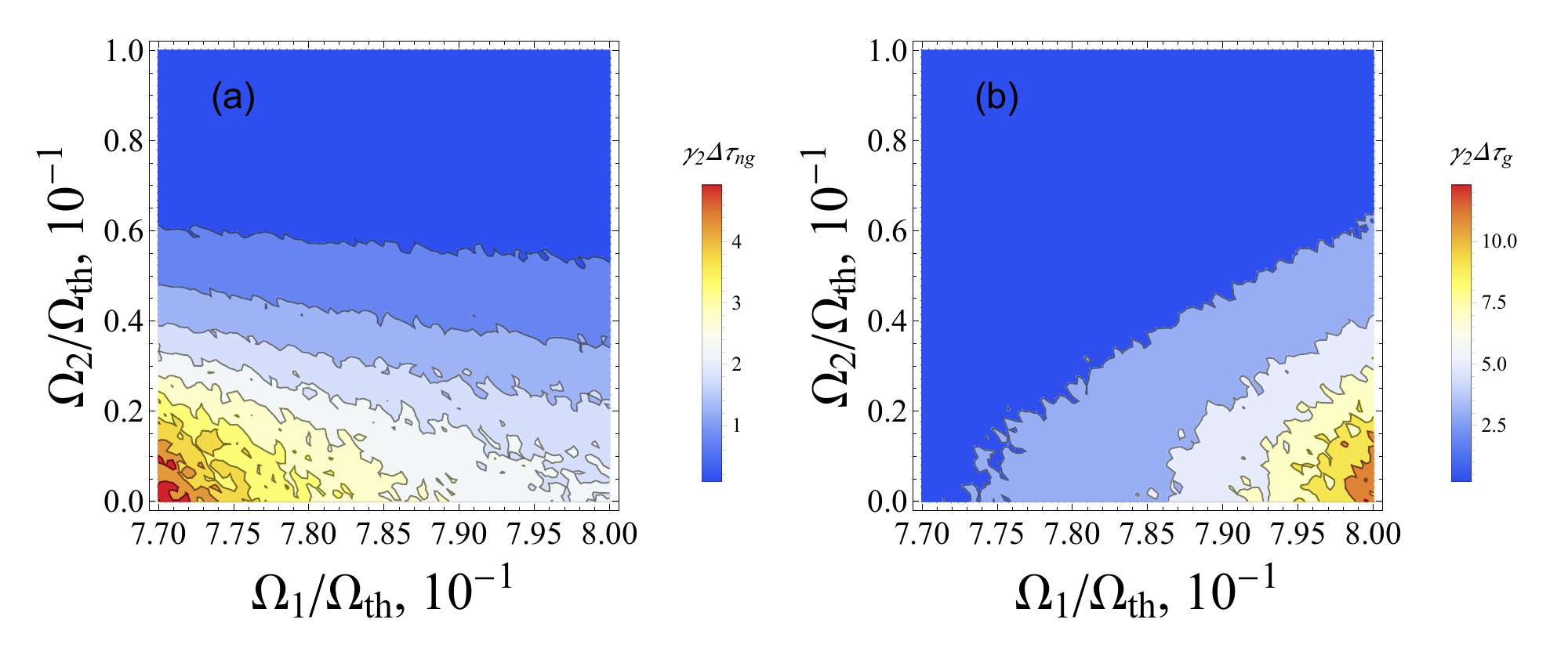}
\caption{The dependence of the standard error of the average lifetimes of the non-generating state, $\Delta\tau_{ng}$, (a) and the generating, $\Delta\tau_{g}$, (b) on the pump amplitude $\Omega_{1}$ and $\Omega_2$. The parameters are the same as in Fig.~\ref{fig:2}.}
\label{fig:B2}
\end{figure}

From the Fig.~\ref{fig:B2} the error for the non-generating state is in the range $0 < \Delta\tau_{ng} < 4.9/\gamma_2$  and the generating state is in the range $0 < \Delta\tau_{g} < 12.3/\gamma_2$.

\section{NIST Verification Test}
\label{app3}

\begin{figure}[t]
\centering\includegraphics[width=0.7\linewidth]{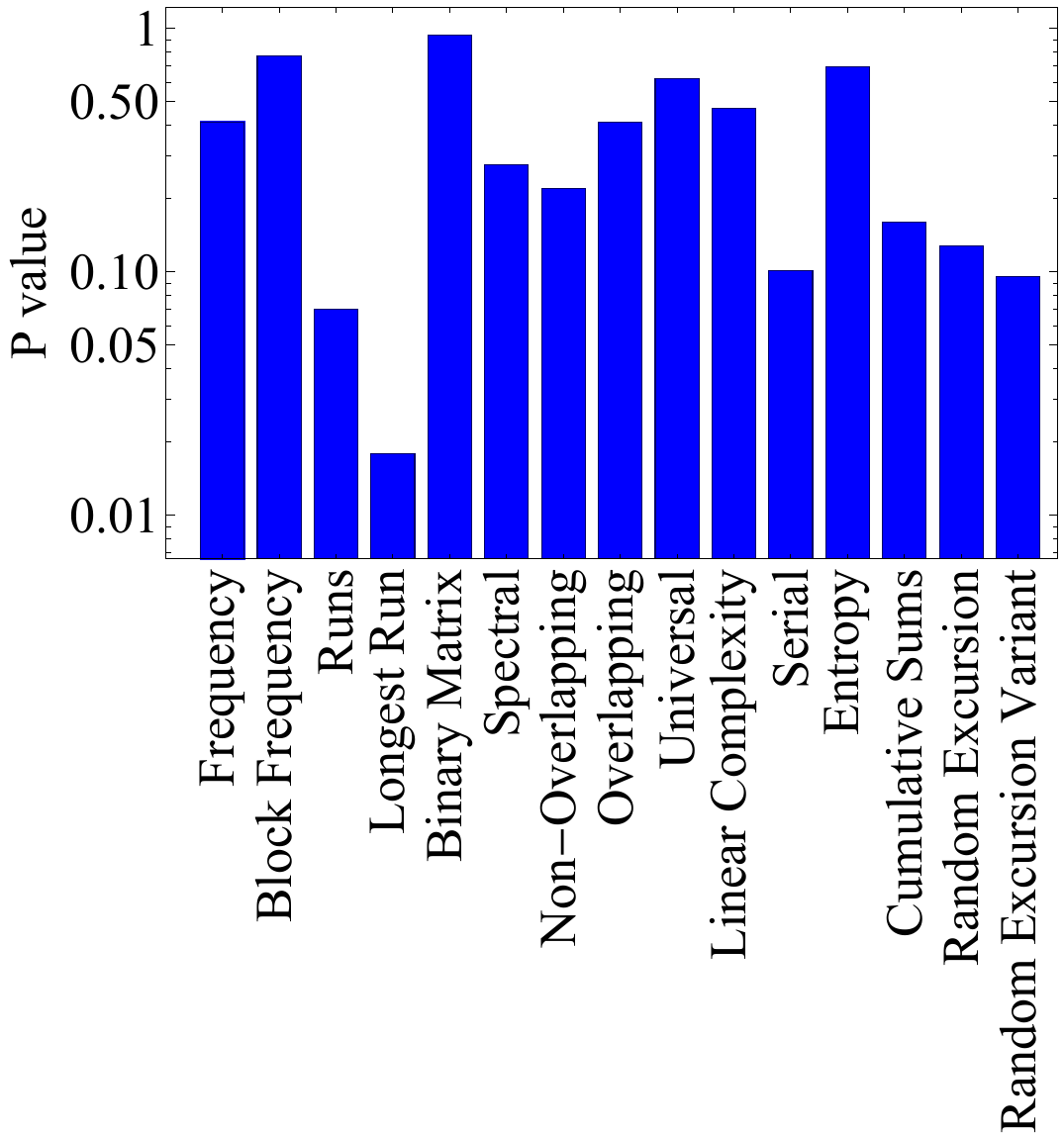}
\caption{NIST test results of the sample 396.308 bits. For the tests with multiple P-value shown the worst one.}
\label{fig:7}
\end{figure}

To pass all NIST tests we generate a sample of 396.308 bits using numerical simulation of stochastic Eqns.~(\ref{eq:2})-(\ref{eq:4}). The simulation parameters are the same as in Fig.~\ref{fig:2} except the pump amplitude $\Omega_1 = 7.8294 \cdot \Omega_{th}$. Each NIST statistical test characterized by the P-value which correspond to the probability that a RNG would have produced a sample less random than the sample that was tested \cite{NIST}. If the P-value is equal zero than the sample is completely non-random, if the P-value is equal one than the sample is absolutely random. To complete the test we use NISTs' recommendation to take a significance level $\alpha = 0.01$, when P-value $> 0.01$ the test is passed. Some tests required a minimal bit sample to pass a test (e.g., Mauerer's Universal Test \cite{NIST}), which required generating the corresponding number of bits. Fig. ~\ref{fig:7} demonstrates tests results for the sample of 396.308 bits, for each test the P-value is greater than 0.01, it means that the bit sample passed all 15 NIST statistical test.

\bibliographystyle{elsarticle-num} 
\bibliography{apssamp}
\end{document}